\documentclass[%
 reprint,
 amsmath,amssymb,
 aps,
floatfix,
]{revtex4-1}

\usepackage{graphicx}
\usepackage{dcolumn}
\usepackage{bm}
\usepackage{xcolor}
\usepackage{ulem}
\usepackage[T1]{fontenc}	
\usepackage[utf8x]{inputenc}

\usepackage{geometry}		
\usepackage{graphicx}
\usepackage[above,below]{placeins}	
\usepackage{times}
\usepackage{float}
\usepackage{natbib}
\begin{document}

\title{Investigating how students collaborate to generate physics problems through structured tasks}

\author{Javier Pulgar}
\altaffiliation{Departamento de Física, Universidad del Bío Bío,, Avda Collao 1202, Concepción, Chile. \\ jpulgar@ubiobio.cl}
\affiliation{Departamento de Física, Universidad del Bío Bío, Concepción, Chile.}

\author{Alexis Spina}
\altaffiliation{Gevirtz Graduate School of Education, UC Santa Barbara, 93106-9490, CA, USA. \\adspina@ucsb.edu}
\affiliation{Gevirtz Graduate School of Education, UC Santa Barbara, CA, USA.}%

\author{Carlos R\'ios}
\altaffiliation{Departamento de Ense\~nanza de las Ciencias B\'asicas, Universidad Cat\'olica del Norte, Larrondo 1281, Coquimbo, Chile. \\ carlos.rios@ucn.cl}
\affiliation{Departamento de Ense\~nanza de las Ciencias B\'asicas, Universidad Cat\'olica del Norte, Coquimbo, Chile.}

{
}%

\date{\today}

\begin{abstract}
Traditionally, scholars in physics education research pay attention to students solving well-structured learning activities, which provide restricted room for collaboration and idea-generation due to their close-ended nature. In order to encourage the socialization of information among group members, we utilized a real-world problem where students were asked to generate a well-structured physics task, and investigated how student groups collaborated to create physics problems for younger students at an introductory physics course at a university in northern Chile. Data collection consists of audio recording the group discussions while they were collaborating to develop their physics problems as well as the solutions they created to their problems. Through interviews, we accessed students' perceptions on the task and its challenges. Results suggest that generating problems is an opportunity for students to propose ideas and make decisions regarding the goals of the problem, concepts and procedures, contextual details and magnitudes and units to introduce in their generated problems. In addition, we found evidence that groups tested the validity of their creations by engaging in strategies often observed with algebra-based physics problems, such as mathematical procedures and qualitative descriptions of the physics embedded in the problem, yet groups invested more time with algebra-based strategies compared to more qualitative descriptions. Students valued the open-ended nature of the task and recognized its benefits in utilizing physics ideas into context, which in turn enabled collaboration in a way not experienced with traditional algebra-based problems. These findings support the use of generative activities as a pathway for students to engage in real-world physics problems that allow for a range and variety of collective processes and ideas.

\end{abstract}

\maketitle


\section{\label{sec:level1}Introduction}
There is a strong tradition among physics education research (PER) scholars to focus on students engaging in well-structured learning activities, such as algebraic problems. However, research has shown the limitations of such activities in fostering conceptual development \cite{Kim,Byun}, and for effective collaboration \cite{HellerHollabaugh,HellerKeith}. Such learning problems tend to benefit from individualized performance \citep{Steiner1966,Pulgar19} rather than collaboration, because of their embedded low levels of positive inter-dependency \cite{JohnsonJohnson,Jonassen}. This motivated us to reflect on alternative tasks that would encourage the socialization of information for collective learning, and would enable processes associated with idea-generation. For such purposes, this work is grounded in the use of a real-world activity consisting of students generating physics problems for younger students. Real-world problems are defined as open-ended activities that demand higher levels of creative thinking compared to algebra-based tasks \citep{HellerHollabaugh,Fortus}. Designing physics curriculum with more creative tasks responds directly to the need of developing appropriate skills for collaboration and innovation for contemporary life and work \cite{Bao2019,sawyerEduInnovation,pllegrino}. 
In this paper, we explore this collaborative work, and physics related ideas and processes students groups engaged in when generating a physics learning activity for high-school students (i.e., real-world problem). We showed the challenges and benefits for using this type of task in physics courses as a way of fostering idea-generation and problem solving strategies often associated with both novice and expert problem solvers \citep{Docktor2015,Dufresne,Gaigher,HellerKeith}. In addition, we explored how students perceived the benefits and difficulties linked to building up ideas with their teammates, and their forms of collaboration with classmates when facing this creative task.

\section{Algebra-Based Physics Problems versus Generating Physics Problems}

Algebra-based physics problems mirror well-defined systems of equations that lead solvers towards unique solutions, \cite{Jonassen}, and are often labeled as textbook physics problems \cite{Chi,Byun,Kim}. Such problems tend to present simplified versions of reality, and are primarily appropriate for the implementation of mathematical representations of physics concepts and principles. Their well-defined nature and purpose makes most of these problems appropriate for the analysis of physics ideas, while an exemption of them might demand students knowledge utilization, the highest level of cognitive demand according to the taxonomy of introductory physics problems \cite{teo}. Differently, qualitative physics problems, also well-defined, have shown richer opportunities for learning \cite{Shing, Docktor2015,Leonard} as well as collaboration \cite{Buteler2016,Leinonen}. 

Studies have found that solving algebra-based problems does not necessarily add to adequate conceptual development \cite{Byun,Kim}. According to Byun and Lee \cite{Byun,Kim}, the best predictor of conceptual development depends on the strategies used for solving the problems rather than the number of problems. Experts problem solvers tend to tackle problems through \textit{knowledge-development}; that is, they begin by building an extensive conceptual understanding before attempting to solve problems \cite{Byun}. Alternatively, \textit{Means-End Strategy} consists of focusing on the problem's goals before attempting to build conceptual meanings, where students work backwards and overlook the meanings associated to solving the activity \cite{Larkin, HellerHollabaugh}. Consequently, algebra-based problems tend to push students towards Means-End Strategy, or `plug-and-chug', known as the practice of finding the formulae that best fit the problem, a behavior typically associated with novice students \citep{Dufresne,Byun,Kim}. The equation-driven approach enacted by novices has also been labeled as bottom-up logic \citep{Meltzer,Shing}, whereas experts are more likely to engage with top-down strategies, where they begin from general principles and then move down to mathematical representations and equations that would enable access to the solution \citep{Dufresne,Larkin}. 

Generating physics problems is associated with real-world tasks \citep{Fortus,Mestre}, which are open-ended and lack constraining conditions (e.g., initial conditions) \citep{Shin,Jonassen}. A real-world problem requires solvers to have the ability to generate subjective assumptions over issues relevant to the problem context, in order to transform and constrain the open-ended scenario into a well-defined one \citep{Fortus,Rietman64}. Consequently, real-world problems are designed for the highest level of cognitive demand, known as Knowledge Utilization \cite{teo,Marzano}, which includes competencies associated with idea-generation and decision making, a set of processes that benefit from collaboration \cite{Dreu2011,Sawyer2009}. For instance, when generating a real-world physics problem, solvers are encouraged to transfer knowledge and make decisions over multiple domains \cite{Mestre}. 
In physics education, algebra-based problems might be rather familiar for students, whom could support their work by recalling past experiences. Nonetheless, generating physics problems as a real-world task, is unfamiliar for students in traditional physics courses. Having been exposed to such cognitive demand is what separates novices and experts from successfully solving real-world problems \cite{Fortus}. According to Fortus \cite{Fortus}, when solving real-world physics problems, the hardest assumptions to make are associated with the absolute or relative magnitudes of the variables involved, a practice rarely found in introductory courses and related tasks. In contrast, easier assumptions relate to physics variables and principles involved in the problems \cite{Fortus}.

The uniqueness of generating physics problems as a real-world activity is found in both the lack of constraining conditions and the great number of features that solvers must decide on (e.g., context, variables, questions, etc.). For instance, Hardy and colleagues \cite{hardy} designed a course where students periodically were asked to create their own multiple-choice questions, resulting in positive learning outcomes, especially for low and middle performance students. According to Mestre \cite{Mestre}, generating questions is a cognitively demanding task that might shed light on the development of students' expertise and knowledge transfer.

\section{Collaboration and Problem Solving}
In the context of problem solving as a group activity, algebra-based physics problems might be associated with disjunctive activities \cite{Steiner1966}, as these do not necessarily demand collective efforts for finding the right solution. Contrary, the activity of generating physics problems might be associated with additive tasks \cite{Steiner1966}, because a good performance would likely emerge as the sum of all members’ contributions and relevant abilities. This creative problem would enjoy higher levels of positive inter-dependency \cite{JohnsonJohnson} compared to algebra-based tasks, as its solution would benefit from collective contributions \citep{Sawyer2003, Baruah,Thompson,Johnson,McMahon}.

For the mentioned reasons, it might not be surprising that both algebra-based problems and the activity of generating a physics problem would respond differently to collaboration. For instance, when facing context-rich problems in mechanics, groups provided better problem solutions than individuals working alone \cite{HellerHollabaugh,HellerKeith}. Recent evidence has shown performance on algebra-based problems is negatively affected by having more social connections in the classroom, particularly if these social ties were used for accessing information \cite{Pulgar19}, however beneficial for generating physics problems. Yet, this effect would strongly depend on whether the learning environment fosters collaboration, as well as motivates students to engage in the processes of idea-generation and decision-making \cite{PulgarPERC}. 

The literature on social networks highlights the importance of social interactions for solving problems \cite{BorgattiCross,Dawson,Grunspan,Liccardi, Putnik}. From here, it is possible to identify two distinctive collaborative mechanisms that would enable group performance: (a) creative combinations, where good solutions emerge when conventional knowledge is combined in original and appropriates ways \cite{Burt2004}; or alternatively, (b) interrogation logic, that is, through a a deep examination of the local knowledge (i.e., physics content) related to the problem \cite{Rhee}. Creative combinations is likely to occur when individuals access novel information outside their groups, whereas interrogation logic tends to be observed in cohesive groups with reduced outside interactions, and whose members invest most of their time and energy in addressing the information managed by their working unite \cite{Pulgar19,Rhee}. Creating physics problems in groups may facilitate either of the latter processes, as team members may feel the need to search for ideas on other groups, in order to access new information to recombine with their unique knowledge \cite{Fleming,Burt2016}. Conversely, because each group must come up with a unique solution, students may feel like it is wasteful to seek out new ideas from other groups, and therefore, collaboration may be observed only within the group. Knowing whether students perceived either of the aforementioned processes as useful for generating problems would shed light on pedagogical mechanism to encourage effective collaboration in physics classrooms and through non-traditional tasks. 

\section{Methods}

In this study, we explore how student groups work to generate a physics problems for younger students. In doing so, we attempted to answer the following research questions: 
\begin{itemize}
    \item a. What are the physics and task related ideas that student groups addressed when generating a physics problem for high school students? 
    \item b. What are the perceived the benefits and challenges of generating a physics problem for high school students?
    \item c. How do student groups collaborate when generating a physics problem for high school students?   
\end{itemize}

\begin{figure*}[htp]
\centering
  \includegraphics[width=1\textwidth]{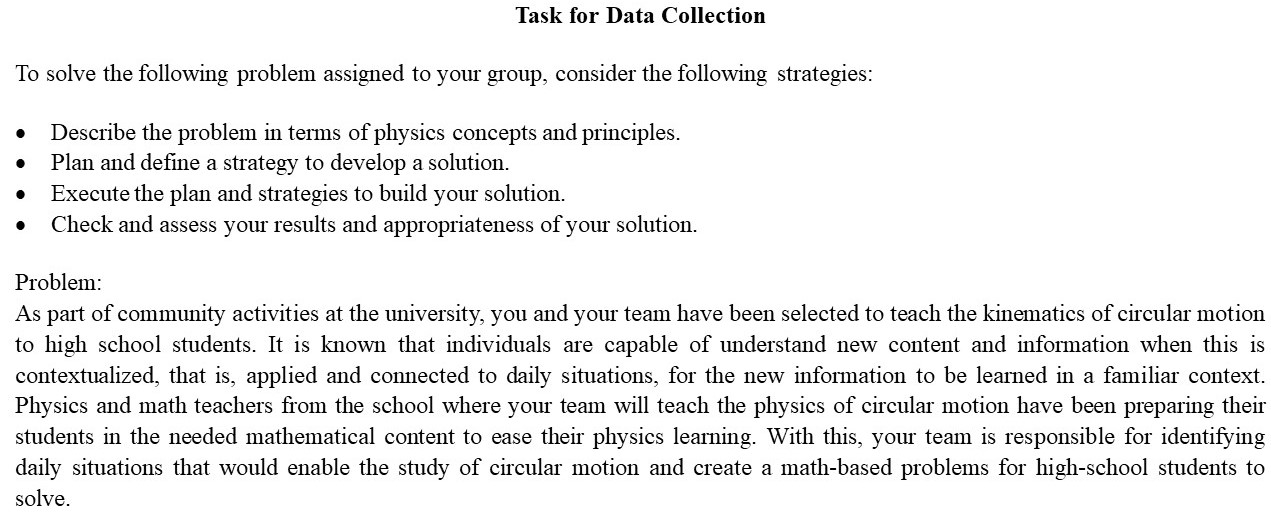}  \caption{Ill-structured problems administered to course participants. \label{Ill-structured}}
\end{figure*}

For this study we observed four student groups from two different sections of an undergraduate physics courses in a University in Northern Chile. The course content consisted of Newtonian Mechanics, which addressed content such as Vector Algebra, Kinematics, Newton’s Laws, Rotational Dynamics and Newton's Law of Universal Gravitation, and participants were engineering majors in their first or second year of higher education. Groups 1 and 2 belonged to section 1, while groups 2 and 3 were observed in section 2. During a problem solving session on the seventh week of the semester, we tasked students with the activity depicted on Fig. \ref{Ill-structured}. 

\subsection{Data Collection and Analysis}
We gathered audio on four student groups during the problem solving session (1.5. hours) on the seventh week of the semester, while they solved the task assigned (Fig. \ref{Ill-structured}). A total of 295 minutes of audio were transcribed by identifying first turns of speech (e.g., \textit{Student 1}: How did you obtain the number of revolutions? Did you multiply the number by something?; \textit{Student 2}: There is one revolution and two revolutions). Later, we revisited the data to separate these turns of speech by message units (\textit{Student 1}: [``How did you obtain the number of revolutions?] [Did you multiply the number by something?]; \textit{Student 2}: [There is one revolution and two revolutions]), as the former may include more than one message unit, and consequently, a variety of ideas may be expressed during the same turn of speech. After the session, we interviewed 4 students (one from each group) and asked them about their experience on creating a physics problem, benefits, challenges and the way they collaborate with their teammates and others. Interviews lasted between 15-20 minutes, were latter transcribed.   

Coding was conducted in NVivo 12 plus software and was intended to elicit the different set of ideas and processes students groups engaged with for generating a physics problem using the concepts and principles of circular motion. First, we reviewed 25\% of the data, and identified emergent issues and ideas students discussed for solving the problem by attending to the dimensions that require decision making, and strategies for solving physics problems (i.e., algebra, conceptual understanding). This analysis led to a first draft of coding definitions for themes. We met with a trained graduate students in qualitative research whose first language is Spanish (research subjects are Spanish native speakers) to review and re-define the codebook for the analysis of the data, with examples taken from the 25\% of data utilized initially, until agreement was reached. Later, we coded 15 (6.25\%) minutes of the transcribed data while negotiating the selection of codes. Finally, the trained graduate students coded independently 45 min (18.25\%) of the data, obtaining a Cohen's Kappa of 0.94 for inter-rater reliability. Finally, we analyzed students' interviews by paying attention on two key ideas: 1. Perceived benefits and challenges of generating a problem; and 2. Nature of the collaboration among students for generating a physics problem. 

\section{Results}

The analysis produced to main themes that emerged during the generation of a physics problem. Table \ref{Tabla1} shows these categories labeled Decision Making and Problem Solving Strategies, and the different themes corresponding to each category. These categories and themes responded to the different sets of ideas, arguments and processes groups engaged in during the activity, which enabled their decision-making and the transfer of physics content into real-world scenarios. First, we describe the nature of these categories and themes, utilizing direct examples taken from the data. Further, we explored students' perceptions regarding the benefits and challenges of the task, as well as the ways in which they collaborate for doing so. 

\begin{table*}[t!]
\begin{center}
  \caption{Codebook of emergent categories and themes addressed by 4 student groups during the task of generating a physics problem. \label{Tabla1}}
 \begin{tabular}{p{3.5cm} p{7.5cm} p{6cm}}
\hline\hline
Code                                                       & Description & Example \\
\hline 
\textit{Decision Making}   &              &       \\

\hspace{0.2cm} Learning Goals                              & Team discusses and makes decisions regarding the learning goals for the generated problem, and the expectation of what the targeted students should learn from it, which mediates the degree of difficulty taking in consideration the school level of the targeted students. 
& \textit{What is the goal of this (activity)? I mean, of teaching this?};
\textit{It would be like explaining them (high school students) how these (movements) work.} \\

\hspace{0.2cm} Physics Concepts \& Procedures             & Team identifies, poses and decides on the physics concepts to use into the problem, as well as the ways in which these concepts align with the generated problem to be well-structured, and consistent with the task requirements. 
&  \textit{I supposed we need to include the equations. That way they only need to replace}; \textit{So the problem must be in order. First you calculate one (value), which then allow you to find other.} \\

\hspace{0.2cm} Problem Context \& Wording                & Team poses and decides on the contextualization of the problem (i.e., place, subjects, actions etc.), and the wording of the problem. 
& \textit{Let us do something cool, like a wooden spinning top.}; \textit{If I want to say  that the car wants to move from A to B, is that displacement?}\\

\hspace{0.2cm} Discussing Magnitudes \& Units            & Team discusses and decides the magnitudes scores and values, as well as measurement units for the physics concepts (e.g., 10 km/h, 20 s, 2.5 km) to be introduced into the problem’s description. 
& \textit{How much do we say the acceleration will be?}; \textit{Do you want the car to get to its destination fast or slow?}  \\

\textit{Prob. Solving Strategies}                         &                &   \\

\hspace{0.2cm} Algebraic Procedures                       & Team describes algebraic steps to obtain physical quantities as a way of solving the problem, normally mentioned to justify the appropriateness of the designed problem. 
& \textit{Because you have the angular speed at 3s, which is 10$\pi$, so 10$\pi$ is equal to (angular)acceleration plus the initial angular speed. So then you clear and get the (angular) acceleration.}\\

\hspace{0.2cm} Physics of Circular Motion in Context     & Team engages in a qualitative description of the physics regarding the circular motion in the context under consideration for the problem. 
& \textit{There is also velocity, this velocity that goes to the middle. This is the one  that enables... This was related to forces if I remember. The topic of the two forces pointing out to one side.} \\
\hline
\end{tabular}
\end{center}
\end{table*}

\begin{table}
  \caption{Frequencies and percentage of categories and themes addressed by 4 student groups during the task of generating a physics problem for high school students. \label{Tabla2}}
 \begin{tabular}{ll}
\hline\hline
Categories and Themes                                & Frequency (\%) \\
\hline
\textit{Decision Making}                             &\\ 
\hspace{0.2cm}Learning Goals	                     & 25 (4.18\%) \\
\hspace{0.2cm}Physics Concepts \& Procedures	     & 209 (34.95\%)\\
\hspace{0.2cm}Problem Context \& Wording	             & 128 (21.4\%)\\
\hspace{0.2cm}Discussing Magnitudes and Units	     & 90 (15.05\%)\\
\textit{Problem Solving Strategies}	                 & \\
\hspace{0.2cm}Algebraic Procedures	                 & 98 (16.39\%) \\
\hspace{0.2cm}Physics of Circular motion in Context	 & 48 (8.03\%) \\
\hline
Total                                               &	598 (100\%)\\
\hline
\\
\end{tabular} 
\end{table}

\subsection{Decision Making}
 The first category (Decision Making) refers to processes related to making decisions and generating ideas in order to create the physics problem. During these processes, teams discussed and decided on issues, such as the learning goals of the activity, the concepts and procedures, and the contextual details of how these elements would be presented in written form. They also decided on the data to be included to make the problem well-defined and the questions that would be appropriate to ask, while also taking into consideration the decided data. From Table \ref{Tabla2}, readers might notice that groups spent most of their time addressing themes related to Decision Making, which included approximately 75\% of the message units analyzed.

\subsubsection{Learning Goals}
Student groups discussed and made decisions regarding the learning goals for their problem, and expectations of what the targeted students should learn. The learning goals of the problem consisted of enabling secondary school students to utilize their physics knowledge (e.g., ``The idea is that they would exercise with the problem we give them''). The school level of the targeted secondary students emerged in multiple opportunities in favor (e.g., ``If we think on the students’ age, they should know how to do such operations.'') or against (e.g., ``Maybe that is too much for a kid in 10th grade. Because that is something they would do in 11th grade.'') the difficulty of the problem in terms of mathematics representations and concepts. In Group 2, this discussion concentrated on the reasons for using circular motion as the key concept, its importance for them (university students) and the targeted secondary students. From the example below, it is possible to perceive the intention of making sense of the activity and the overall objective of learning concepts and principles of circular motion: 
\begin{quote}

Student A: What is the goal of this (activity)? we mean, of teaching this? 

Student B: What thing? 

Student A: All this equations and concepts.

Student B: For us or for students who would be solving this?

Student A: Well, for both. 

Student B: It is assumed that almost every movement is circular, as there is rare to find truly straight movements. These do not exist. You will see that this (movement) has no angles, but in reality it has.

Student A: It would be like explaining to them (high school students) how these (movements) work.
\end{quote} 

The discussion about the learning goal emerged from the explicit objective of the task, yet was not necessarily supported by the argument made by student B in regards to the ever-present circular motion, which attempted to highlight its importance by transferring and extending the real nature of motion upon a combination of circular displacements. The latter argument is interesting, as students explicitly attempted to make sense of the content to be learned for the targeted students, yet this idea received no follow up from other team members. 

Group 2 linked the goal of the task to what science teachers would do in the face of a similar task. The following segment shows this brief moment of reflection, where the group attempted to convey the appropriateness of their problem in coherence with the learning objectives:    
\begin{quote}
   Student C: We have to be clear with the goal, which consists of teaching and learning kinematics of circular motion to 12th grade students. So, it is like a teacher preparing to teach circular motion. That way, each element of circular motion could be linked to different contexts from daily life, or just one. It has to be didactic for them (high school students) to understand.
   
Student D: Let us do problems similar to the ones our instructor has used.

Student C: That is it, didactic.
\end{quote}
 
This segment provided evidence that, in finding the learning goal, students mirrored what experts [secondary school teachers] would do in contextualizing the content, and projected their own expectations into what a physics problem should look like. Finally, in Group 3, we observed a deeper reflection of the learning goal, where a student highlighted the importance of the real-life context for learning:   

\begin{quote}
    So, if you include a difficult exercise, but they do not know how to solve it through equations, they would remember that they tackle a problem involving a laundry machine where there was a circular motion and were able to calculate the speed. Consequently, and lastly, they would understand and know how to calculate angular and tangential speed for a laundry machine, and they would imagine the same type of motion but on different problems.
\end{quote}

	According to this quote, the context would mediate the difficulty of the problem in the case that the targeted students were incapable of using the needed equations, as they would ultimately associate the context of the laundry machine with circular motion. Through this link, the student argued that learners would draw similarities in the use of equations for the purpose of calculating quantities across different scenarios. This, in essence, is the notion of transferring knowledge, or in other words, the use of information from a well-known to an unknown situation.  

\subsubsection{Physics Concepts and Procedures}
\begin{figure*}[htp]
\centering
\includegraphics[width=1\textwidth]{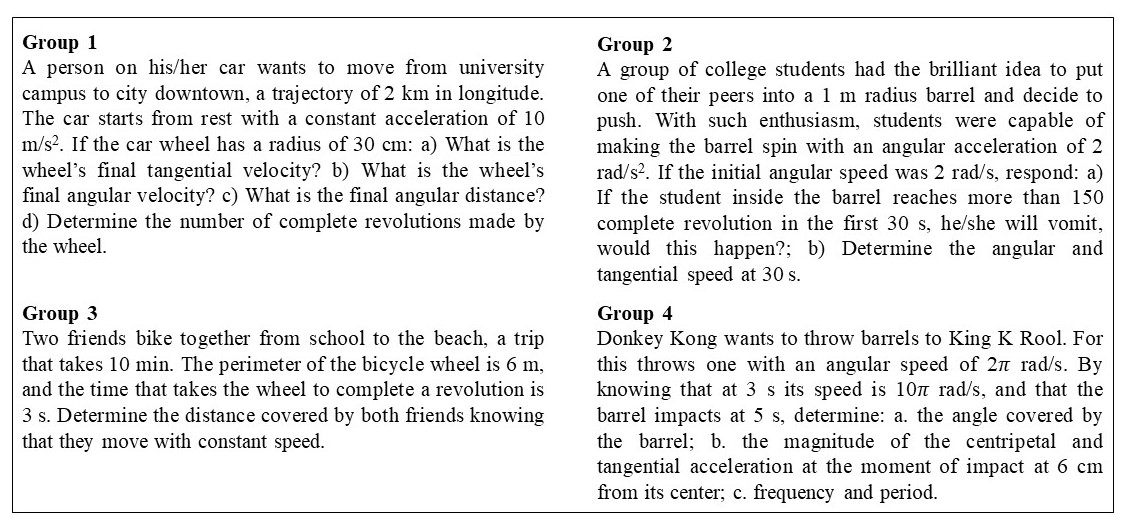}  \caption{Problems generated by groups 1, 2, 3 and 4.\label{GroupProblems}}
\end{figure*}

During this process, teams identified, posed, and decided on the physics concepts to use in the problem. In addition, they looked at how these concepts aligned for the generated problem to be well-structured, and consistent with task requirements. Defining procedures was in direct connection with the learning goal, as teams engaged in the former process to meet the expectations previously defined. When addressing this theme, groups attended to and emphasized different sets of elements, such as the algebraic steps through manipulation of equations, concepts and the combination of quantities for an appropriate problem structure. Figure \ref{GroupProblems} would allow readers to identify the set of concepts used by each group on their respective problems. For instance, Groups 1 and 2 from section 1 decided to use the angular version of speed, distance, acceleration as data to determine the magnitudes defined as questions. In contrast, Groups 3 and 4 from section 2 selected the linear version of speed, distance and acceleration as initial conditions that would allow solvers to determine the number of revolutions completed at the end of motion, the final distance covered, speed, and other questions. 

We observed two different sets of strategies for addressing the early stages of this process and making decisions: Equation-driven and Concept-driven. The first approach (Equation-driven) was observed in teams that primarily focused on the mathematical dimension of the problem for decision making, whereas concept-driven strategies emphasized the conceptual dimension of the situation to then reflect on mathematical representations. Group 1 mostly utilized the Equation-driven approach when defining concepts and procedures. In doing this, they proposed to include the equations in the problem so that students could easily ‘plug and chug’ and find the solution (e.g., ``I supposed we need to include the equations. That way they only need to replace what's there.''). Consequently, this process was guided by the (implicit) idea that a problem is constructed in the same way that one may solve it, which refers to following very structured set of steps (e.g., ``So the problem must be in order. First you calculate one (value), which then allows you to find other.''). The example illustrates the algorithmic nature of developing a problem for Group 1. Similarly, Group 4 engaged in such a strategy for defining concepts and procedures, yet transitioned towards a Concept-driven description of the phenomena after establishing the situation to be used: 

\begin{quote}
 It will start from rest, and that way we could calculate the movement of the barrel. So then, we would tell them that the barrel is accelerating constantly and that it needs certain time in seconds to hit the target. Because after some seconds the barrel will be there, at its final position. That is, it will impact the target, so then they could begin their calculations for different things, like angle and everything.
\end{quote}

This quote from a member of Group 4 showed a simple physics analysis of the situation used (i.e., a barrel is thrown to a person), as the student analyzed the position and evolution of the object through time, and enabled further understanding of the procedures the targeted students are expected to go through for solving the problem.

Even though deciding on concepts and procedures utilizing Concept-driven approach is not absent from the attention to equations for decision making, the subtle difference is that the equations emerged after deciding on the concepts first. For instance, a student in Group 2 stated: ``So let's create a situation where we combine angular speed, acceleration and everything else, like a situation that involves circular distance.'' This approach helped the group to create a problem based on the relationship between concepts rather than on the exclusive use of equations. In Group 3, this was insinuated by a member arguing against the equation-driven approach in the description of the physics regarding the situation selected: 
\begin{quote}
    
More than the equations, it would be better to say the there is a force acting over there, whereas there is another force but in that direction… We need to be more specific. For instance, say that there is a force acting to the inside, and another to the outside.
\end{quote}

 Even though the use of forces is beyond what is expected for the target students to know, this qualitative description provided a conceptual framework for the group to decide on the physics for the problem.

\subsubsection{Physics Context and Wording}
Groups posed ideas and decided on the contextualization (i.e., place, subjects, actions etc.) and wording of the problem. This process required groups to invest considerable amount of time (21,4\% of the data, see Table \ref{Tabla2}), which is not surprising taking into account the need to select a daily situation  where circular motion is observed. Groups experienced some conflict to find the right contextual elements to use. For instance, Group 1 started by only focusing on objects with wheels, yet members attempted to achieve some degree of novelty by pushing the conversation towards situations beyond wheels (i.e., ``Is there one without wheels?  I cannot think of anything.''). Group 4 wanted to create something fun for students to be motivated with (e.g., ``Let us do something cool, like a wooden spinning top.''). Other participants suggested ideas like a fisherman moving his fishing rod and describing a circular motion, or an ant walking on the inner wall of a bottle. Here, originality was controlled by their level of confidence in exploring such situations through their physics knowledge, (e.g., ``We do not need to complicate ourselves with that.``) and ended up using more familiar situations. 

As seeing in Figure \ref{GroupProblems}, Group 1 selected the wheels of a car moving and covering a trip between places; Group 2 decided on the use of a person trapped in a barrel in motion; Group 3 decided on the motion of a bicycle; and Group 4 utilized references from a problem created by their instructor to design a situation where Donkey Kong moves a barrel. 

The wording of the problems illustrated the use of technical language, and became an opportunity for participants to challenge their own conceptual understandings. A simple example emerged from Group 1, where a student asked the following when deciding on the right wording of the problem: ``If I want to say that the car wants to move from A to B, is that displacement? `` Because displacement is defined as a vector, then to properly use it, the group needs to incorporate a direction. There is no evidence in the data when this correction was made, but the problem is worded using scalars rather than vectors, and framed the phenomena as ``a car wants to move''. 

Another example of negotiating the wording with a correct use of physics concepts was observed in group 4, when students discussed the conditions under which Donkey Kong would make the barrels move: 
	
\begin{quote}	
Student P: For this, he tosses the barrel from rest?

Student O: You cannot toss a barrel from rest. It releases the barrel then. He let the barrel go.

Student P: Ah, okay.  

Student O: Or better, he tosses it with an initial speed, is that okay?   
\end{quote}

This segment showed students’ understandings of motion in connection with an appropriate use of language to convey the idea that releasing and tossing the barrel implies different physical conditions. Here, a body that begins its motion from rest must be release to accelerate due to the presence of an external force (e.g., gravity), and will therefore gain speed. Differently, tossing implies an interaction (i.e., force) that boosted energy and therefore speed to the object that was moving. Both ideas showed comprehension of motion and the implications of forces for the motion defined in the problem.

\subsubsection{Discussing Magnitudes and Units}
During this process, groups engaged in discussions and decision making regarding relative values and magnitudes, as well as measurement units for the physics concepts (e.g., 10 km/h, 20 s, 2.5 km) to be introduced into the problem’s description. This process is important as it brings a sense of ‘reality’ to the physics of the phenomena and situation under design. This process enabled groups to identify and select appropriate values to link with the physics concepts used as data. The validity of these magnitudes was tested through the calculation of their unique responses. For instance, Group 2 discussed the appropriateness of a high angular acceleration for the barrel that yield to 1,400 rpm (revolutions per minute), and decided to `Lower the values’. Similarly, in Group 1, we observed the following interaction for deciding on the acceleration of the car: 

\begin{quote}
Student A: What do we say the acceleration is?

Student B: 20. 

Student A: 20 what? 

Student B: Meters by square second.

Student A: Is that too much?

Student B: I know it is a lot. Do you want the car to get to its destination fast or slow? 

Student A: I want it to get there at a normal speed. 
\end{quote}

	This dialogue reflects the intention to utilize magnitudes that resemble real life situations. Later on in this process, the same group tested the problem with an acceleration of 10 m/s, and obtained a final speed of 200 m/s, an unrealistic result for a common car moving in the city. 
	
\subsection{Problem Solving Strategies}
The following set of themes emerged from students engaging in processes often associated with solving algebra-based problems, where students are likely to utilize physics concepts, their mathematical representations, and enact on physics descriptions connected to the context of the activity. The literature on novice and experts physics problem solvers suggest that the former group tends to utilize algebra-based strategies (e.g., ‘plug-and-chug’) rather than qualitative descriptions, a strategy associated with expert behavior \citep{Shing,Docktor2015}. 

\subsubsection{Algebraic Procedures}

This process relates to algebraic steps that the group went through to obtain physical quantities needed to solve and test the appropriateness of their  designed problems. This included suggesting strategies to determine a physics quantity (e.g., ``Here we will use a proportionality rule. If one revolution is 2$\pi$, then x revolutions will be…''); and/or requesting advice on how to proceed in order to get the right value (e.g., ``How do we transform this to radians? Does someone know how to?''). Most of the evidence found here emerged when students wanted to achieve either of the latter two goals.  

	To contextualize the use of algebra in this context, it is important to remember that kinematics problems rely on three fundamental physical quantities: position [$\vec{r}(t)$], velocity [$\vec{v}(t)$] and acceleration [$\vec{a}(t)$], all functions of time $t$. Even though these concepts are defined as vectors, in this context students utilized these mathematical representations to determine scalar quantities, or the magnitudes of the vectors at any given time. In circular motion, these concepts are written in an angular form: angular position [$\theta (t)$], angular speed [$\omega (t)$] and angular acceleration [$\alpha (t)$]. The link between these linear and angular magnitudes comes from $s=\theta R$ (i.e., arc) , $v=\omega R$ and $a=\alpha R$. Consequently, in order to test their problems, students manipulated some or all of the latter mathematical relations. For instance, students in Group 4 had the following argument to determine the angular position of distance covered by the barrel: 

\begin{quote}
Student L: And how would I get the angle?

Student M: With the (angular) acceleration that is obtained from the equation. With the angular speed. Because you have the angular speed at 3 s, which is 10 $\pi$, so 10 $\pi$ is equal to (angular) acceleration plus the initial angular speed. So then you clear and get the (angular) acceleration.
    
\end{quote}

	Here, student M suggested the use of angular speed [$\omega (t)$] at time 3 s, to determine the value of angular acceleration by isolating this value from the equation, because all other elements were given. Once this was done, the argument, although not explicitly mentioned, oriented to the use of this value of angular acceleration into the equation for angular position, and calculate $\theta$ at 3 s. Once again, this was possible because all other elements in the equation are defined. 
	
	Another interesting example was observed in Group 1, as they used the equations $\vec{r}(t)=\vec{r}_0+\vec{v}_0 t + 1/2\vec{a}t^2$ and $\vec{v}(t)=\vec{v}_0+\vec{a}t$ to determine the time that it will take the car to reach its destination. The following interaction depicts the set of algebraic steps suggested for one member to achieve this goal: 
	
	\begin{quote}
Student A: So we have that the initial position is zero, and the initial speed is zero. And we have that (in the equation), only the acceleration for the square time will remain.   

Student B: And then?

Student A: That will give you 10 m/s (magnitude of the acceleration) times the square time, and with that you can obtain the time. So, it will be the square root of something, and then we will use only the positive root. 
	    
	\end{quote}

	This interaction shows an appropriate use of the equation $\vec{r}(t)=\vec{r}_0+\vec{v}_0 t + 1/2\vec{a}t^2$ to obtain the time, given that all the elements but the time are known (final distance is given in the heading of the problem and is equal to 2 km). Because mathematical manipulation may be perceived as a rather individual exercise, it is not surprising that the audio recording only captured brief descriptions of the strategies to be implemented to find numerical values, and the request for advice on how to calculate them.  

\subsubsection{Physics of Circular Motion in Context}
This theme consisted on groups addressing the qualitative description of the physics regarding the circular motion in the context under consideration for the problem. This process enabled access to students’ conceptualization of the physics phenomena in question and the ways in which they would explain such situations. The sample size of examples that illustrate this process is rather small and does not provide evidence on the most difficult concepts. Consequently, the observed frequency (see Table \ref{Tabla2}) reflects the disparity between algebraic-versus-qualitative strategies students displayed. Moreover, qualitative descriptions emerged from the data when students tried to make sense of the situations and physical objects considered for the problem. 

There is evidence that students attempted to explain revolutions, tangential velocity, and inertia, the last being used to determine what would happen if someone were to fall from a fast-spinning carousel. The concept of a revolution was conceptualized through the perimeter of a wheel: ``Suppose that first there is a point that moves along the perimeter until here. That will be one revolution''. This description is very simple and does not really reflect deep understanding of a physics concept that one were to associate with motion. 

A more interesting example was provided by Group 4 in an attempt to understand the relationship between angular and linear speed in the context of a wheel moving. First, angular speed is defined as the change of angular position per unit of time (i.e., $\omega=\Delta \theta / \Delta t$), and may be difficult to understand because it does not imply distance units, such as meters or kilometers. Secondly, an object spinning will measure the same angular speed at any distance from the center of rotation (i.e., radius) at a particular time. However, the linear or tangential speed will increase according to the distance from the center of rotation as shown by the equation ($v=\omega R$). Next, the discussion unfolded as follow: 
	
	\begin{quote}
Student E:  If this is supposed to be in the same wheel, then why? If you advance five meters, you will complete the same number of revolutions.

Student G: Yes, you are right. So, how many…

Student F: The angular speed will change at different points of the wheel.
	    
	\end{quote}

	As one would notice, the claim made by student F suggested a misconception regarding the nature of angular speed, because this quantity remains constant regardless of the distance from the center of rotating object. Consistent with the definition of angular speed, the comment made by student E would have made more sense if instead of using 5 meters as the distance covered, he would have used angular measurements to highlight the distinction with linear speed. 
	
	The last example that illustrates the nature of qualitative descriptions used by students emerged in Group 3 when discussing the nature of acceleration and forces on circular motion: 
	
	\begin{quote}
Student S: There is also velocity, this velocity that goes to the middle. This is the one that enables… This was related to forces if I remember. The topic of the two forces pointing out to one side. Now I remember, centripetal and centrifugal force.

Student T: Centrifugal force was like…

Student U: It is the one that points inside. 

Student T: No. 

Student S: Centripetal force points to the inside. 

Student T: Okay, centripetal force points to the inside. But centrifugal points to the outside. And those two forces would make that… ``There were like equals and… 

Student U: Both forces allow the circular motion. 

Student T: But this centrifugal force was something like hypothetical, or something that was not real…
\end{quote}

According to the interaction, students attempted to make sense of the physical interactions that enable circular motion: in this instance, centripetal force. The ideas related to centripetal force are correct: it is an interaction that is directed to the center of the circumference described by the motion, and it is responsible for the circular motion. However, centrifugal force, as corrected by student T at the end of the interaction, is not a force but rather the effect of inertia, defined as resistance to change the state of motion and is often referred to as a ``fictitious force''. Finally, this interaction provides some insights on students’ understandings, but again fall short to give substantial evidence to assess whether students are actually understanding the underlying physics of circular motion beyond the use of equations.

\section{Student Experience}

Students shared their perceptions regarding generating a task while taking into consideration their experience with algebra-based physics problems in school and university physics courses. From here, they recognized the differences between addressing such well-structured activities and that of creating a physics problem for younger students. Among the challenges of these creative tasks, student 1 (from Group 1) posed the lack of familiarity with the process of coming up with appropriate assumptions and different ideas for unique solutions. Such difficulty has been documented in Fortus \cite{Fortus}, and identified as a reason why novices and experts performed differently on real-world problems. In line with this, the same participant suggested the learning benefits of using open-ended problems more often, as a way to get accustomed with the requirements of tasks where they need `to create, search and define the problem'. 

For student S2 (from Group 2), the creative nature of the task was perceived as an interesting feature that allowed students to `think in what to do, and to generate the problem that you want.' Student 2 (S2) suggested that even though this activity pushed him to reflect more on the content because it is more engaging, he doubted whether this generative activity would foster higher level learning compared to traditional problems: `I do not know if you learn more compared to a good [well-defined] example with formulae and its application'. For him, the biggest bit of uncertainty was linked to the open-ended nature of the tasks and not mastering the content: `the tendency to include more that what it is actually needed, because you are not sure whether with this or that you are going to get the expected results'. 

For student S3 (from Group 3), generating problems allowed her to use her mind more so than algebra-based problems. According to her, `one gets more by doing things in real-life because that is what you will face at the end [work environment]. For me personally, that it is more challenging, but I liked them more because they take me to what I would do in few more years'. Additionally, student S1 argued that generating a physics problem pushed him beyond the ideal scenario where physics magnitudes are often introduced in the classroom, but in a real context. Further, he (S1) recognized this as a difficult mental exercise due to the lack of constraining conditions, as for a real-world phenomena one must consider `all the variables involved'. Such a challenge may be linked to a lack of familiarity in having the liberty to manipulate and utilize physics magnitudes and other variables, and not knowing whether a magnitude set by his group was `approximate, more or less to what would be real'.

\subsection{Collaboration}
According to students S1,S3 and S4, collaboration for generating a physics problem occurred mostly within the group, where they recognized that their solutions came up after collective decision-making, but without the need to reach out to members of other groups. This collaborative mechanism is associated with interrogation logic \cite{Rhee}, where team members pay attention to the knowledge managed by team members for solving the problem. Interestingly, students claimed that solving an algebra-based problem is a process that motivates social interactions outside their group (e.g., creative combinations), whereas addressing a generative task somehow discouraged such socialization beyond their teams, because their solutions (i.e., generated problems) were different. Further, students agreed that depending on the problem, the nature of the social interaction within and between groups was different in their purpose. 

For instance, when addressing algebra-based problems, student S1 suggested that `the ones [team members] who normally solve the close-ended [algebra-based] problems are always two people. They are always the same two, because for them it is easy to seek out the information to include into the equations, and then get the result...' This statement is coherent with problem solving strategies well documented in PER literature, where students would seek out appropriate information for `plugging \& chugging' in order to get the numerical results to their problems. Additionally, student S1 added: "For close-ended problems, because for everyone it is the same [activity], one search for the method they are using, and if you get lost then someone else may have the answer.' The type of information he claimed to seek out in algebra-based problems refers to `specific knowledge', with little attention on the `why you are using this?', but more on `finding the results and finishing the exercise'. Similarly, student S4 suggested that for solving algebra-based problems, the interactions for information seeking often took the form of `what equation did you use?' or `what did you get?' Contrary, when generating a physics problem, student 4 suggested that the nature of the interactions enabled their team to annotate ideas from the brainstorming process that were later discussed, discarded or agreed upon by all group members for further development. Student S2 added that 'in an open-ended [generating a physics problem] problem we four have to do it, because we all have to pay attention in case something it's left out, because these [problems] require more information.' On this context, for student 1 the search for information takes a more complex form, as 'you need to take what others [team members] are doing into what you are doing', and for such purpose he highlighted the importance of understanding why others perform in the way they did (e.g., `Why you did that?' And `How you did it?'). For student S4, the nature of generating a problem is `more subjective', and therefore discouraged the interaction with other groups in order to check for what they are doing. In the same line, for student S2 there is no need to interact with peers outside the group `because all problems [solutions] are different, and I do not need to compare my response with other person because it will be completely different'. 

\section{\label{sec:level42}Discussion}

  Generating real-world physics problems is an authentic activity for educators to engage in. Creating problems that require students to generate, apply, and select subjective assumptions \citep{Fortus,Rietman64} is complex and requires further exploration in education. Our first area of results, Decision Making, constituted the different dimensions that required subjective assumptions in order to solve the problem. These ideas included: 1. Learning goals; 2. Physics concepts and procedures; 3. Problem context and wording; 4. Magnitudes and units.  The second category of results, Problem Solving Strategies, involved 1. Algebraic procedures and 2. Physics of circular motion. According to Fortus, assumptions regarding the physics variables and principles, and regarding the magnitudes of these variables, are the two main assumptions necessary for solving real-world problems. The first assumption (i.e., physics variables and principles) is easier to make for novices (e.g., undergraduate physicists) and experts (e.g., graduate physicist), compared to assumptions regarding the numerical magnitudes of the variables used in the problem \cite{Fortus}. 

Connecting with the themes from Decision Making, we argue that Physics Concepts and Procedures mirror the first type of physics assumption that is accessible to both novices and experts, whereas Magnitudes and Units might be consistent with the second type of assumption, which more experts are familiar with. Extending the dichotomy, we propose that assumptions about the Problem Context and Wording as an alternative and a more accessible assumption to make for both novices and experts, as both groups of students are likely to have experience reading different types of algebra-based problems, with various contextual details and wording. Therefore, students may be more efficient in using that knowledge as a resource for making their own assumptions. In contrast, and even though all participants have been exposed to learning activities of diverse nature, discussing and making decisions about the problems’ Learning Goals may be more challenging, as this entails knowledge of the target students, which will ultimately mediate the problem’s level of difficulty. In sum, having students generate problems adds two alternative types of assumptions with arguably different levels of complexity for both experienced and non-experienced solvers in Problem Context and Wording, and Learning Goals. 

In addition, results show that developing problems encouraged students to engage in both quantitative (Algebraic Procedures) and qualitative (Physics of Circular Motion) strategies for testing (solving) their problems. It is important to recognize the differences in time invested in these problem solving strategies, where students tended to favor algebraic procedures over qualitative descriptions. Reducing the gap between the time invested in algebraic procedures and qualitative descriptions of the content constitutes an additional challenge for physics educators, as shown in the literature \citep{Byun}, and further pedagogical innovation and research needs to be conducted on this matter. For instance, one may think about using characteristics from isomorphic sets of physics problems \citep{Meltzer,Shing} in order to encourage students to generate problems with such characteristics (i.e., quantitative and conceptual problems around the same content). Here, generating both mathematical and conceptual problems from the same content may increase reflection of the content beyond the utilization of mathematical representations. In the likeliness that students begin by algebraic procedures, one may explore qualitative descriptions. 

Students' perceptions of facing generative tasks sheds some light on some of the benefits and challenges. This creative task was recognized for giving students the chance to utilize their knowledge in a real-world context. Again, the lack of familiarity regarding how to engage in real-world problems was highlighted as one dimension where they need more practice. Consequently, education and physics education might require more opportunities for students to engage in generating assumptions and making decisions based on information, a key set of practices for 21st Century education \cite{pllegrino, Bao2019}. 

Further, the ways in which student groups collaborated for generating the problem shed light on the nature of the social processes students tend to engage in, and the importance of such processes in supporting learning. For instance, social interactions for solving algebra-based problems responded to the `bottom-up' logic \cite{Meltzer, Shing}, in that students pursued  specific information to utilize for solving the problem, rather than enacting on a content-oriented strategy, like `top-down' logic \cite{Dufresne, Larkin}. Even though this process might have resembled the mechanism of creative combinations defined in the social network literature \cite{Burt2004,Burt2005, Rhee}, as students claimed to seek out information to other groups, the nature and, ultimately the purpose and content of the information requested signals a practice that does not aim for creative ideas to emerge, but rather the reproduction of conventional knowledge in the face of algebra-based physics problems. In contrast, generating a physics problem in a traditional physics classroom enabled in-depth group discussions, where students paid attention to the ideas and knowledge managed by their team members (i.e., interrogation logic) \cite{Rhee}, and where students recognized the value of building upon each others' ideas for coming up with a unique solution. The nature of such collective process is evidence of the additive nature of the task \cite{Steiner1966} and its interdependence \cite{JohnsonJohnson} that motivated them to understand procedures and knowledge utilized by others.

\section{\label{sec:level43}Conclusions}
The evidence shown in this study encourages the use of generative physics activities, and more specifically, the task of creating problems. This would allow student groups to spend more time addressing the multiple dimensions of decisions that must be made, and thus prioritizing the process of idea-generation, here engaged through effective forms of collaboration within their work groups. Moreover, the extent to which the percentage of engagement leads to better results in terms of the quality of the problem would depend on the effectiveness of these processes and overall collective performance. However, evidence that students would dedicate significant portions of their time to generate ideas for the creation of problems, or solutions in general, is likely to boost familiarity in the face of idea generation using concepts and principles of the curriculum, which may ease transfer and the development of deep learning. Finally, developing the right set of skills for facing real-world problems seems to be valued not only from the great scheme of education, but also for students. Consequently, more efforts and innovations must be made in order to positively respond to such demands.

\acknowledgments{This material is based upon work supported by the AAPT E. Leonard Jossem International Education Fund. Any opinions, findings, and conclusions or recommendations expressed in this material are those of the author(s) and do not necessarily reflect the views of the American Association of Physics Teachers.

We appreciate the help and advice from Valentina Fahler regarding data analysis. 
C. R. is supported by Vicerrector\'ia de Investigaci\'on y Desarrollo Tecnol\'ogico, Universidad Cat\'olica del Norte, through Grants VRIDT-FPEI 2018.
}

\bibliographystyle{unsrt}
\bibliography{ref}

\end{document}